\documentclass[]{article}

\usepackage{url,hyperref,lineno,microtype,subcaption}
\usepackage[onehalfspacing]{setspace}
\usepackage{amsmath}
\usepackage{graphicx}
\usepackage{pdfpages} 

\begin{document}
	\onecolumn
	
\begin{flushleft}
	{\Large
		\textbf\newline{Topical alignment in online social systems} 
	}
	\newline
	\\
	Felipe Maciel Cardoso\textsuperscript{1,2, *},
	Sandro Meloni\textsuperscript{2,4},
	Andr\'e Santanch\`e\textsuperscript{1},
	Yamir Moreno\textsuperscript{2,3,5}
	\\
	\bigskip
	\textbf{1} Institute of Computing, State University of Campinas, Campinas, 13083-852, Brazil
	\\
	\textbf{2} Institute for Biocomputation and Physics of Complex Systems, University of Zaragoza, Zaragoza, E-50018, Spain
	\\
	\textbf{3} Department of Theoretical Physics, University of Zaragoza, Zaragoza, E-50009, Spain
	\\	
	\textbf{4} IFISC, Institute for Cross-Disciplinary Physics and Complex Systems (CSIC-UIB), 07122 Palma de Mallorca, Spain
	\\
	\textbf{5} ISI Foundation, Turin, I-10126, Italy
	\\
	\bigskip
	
	* fmacielcardoso@gmail.com
\end{flushleft}

\begin{abstract}
	Understanding the dynamics of social interactions is crucial to comprehend
	human behavior. The emergence of online social media has enabled access to
	data regarding people relationships at a large scale. Twitter, specifically, is an
	information oriented network, with users sharing and consuming information. In
	this work, we study whether users tend to be in contact with people interested in
	similar topics, i.e., if they are topically aligned. To do so, we propose an approach based on
	the use of hashtags to extract information topics from Twitter messages and
	model users' interests. Our results show that, on average, users are connected
	with other users similar to them. Furthermore, we show that topical alignment provides
	interesting information that can eventually allow inferring users' connectivity. Our
	work, besides providing a way to assess the topical similarity of users, quantifies
	topical alignment among individuals, contributing to a better understanding of
	how complex social systems are structured.
\end{abstract}

\section{Introduction}
Relationships among people determine the structure of complex social systems. As such, the emergence and widespread use of online social networking sites have allowed to address a number of questions related to how humans connect among each other. Research using data from online social media have, in turn, produced new methods and models that are at the core of present (computational) social sciences. In this work, we explore the relationships between users of the microblogging service Twitter and the information shared by them. Information sharing is a very important aspect of Twitter, which is also considered as an information network \cite{Myers2014a}, i.e., it is often a means for the consumption and sharing of contents that are mainly diffused through users' connections. The way Twitter $-$and other social networks $-$ works leads to an interesting linkage between information and (often adaptive or dynamic) relationships among individuals, which is the focus of our investigation. 

In what follows, we inspect how much the information shared by users is related to their connections in the social network. Our goal is to demonstrate that the information spread in Twitter is a crucial component of social dynamics through the verification of topical alignment. 
Connected users being topically aligned is an indication of how much homogeneity is pervasive across their dimensions of interests and ideas. Accordingly,  this requires a proper assessment of the information topics, since focusing only on individual annotations does not capture the latent ``context'' in which users are engaged when exchanging messages. 
We infer topics from clusters of highly associated \textit{hashtags}  in messages exchanged by users. This allows us to capture topics exposing latent higher-level semantic entities without the need of an external ontology or manual classification step \cite{Lehmann2012, Lee2011, Bogdanov2013}.

\begin{figure}[h!]
	\centering
	\includegraphics[width=0.8\columnwidth]{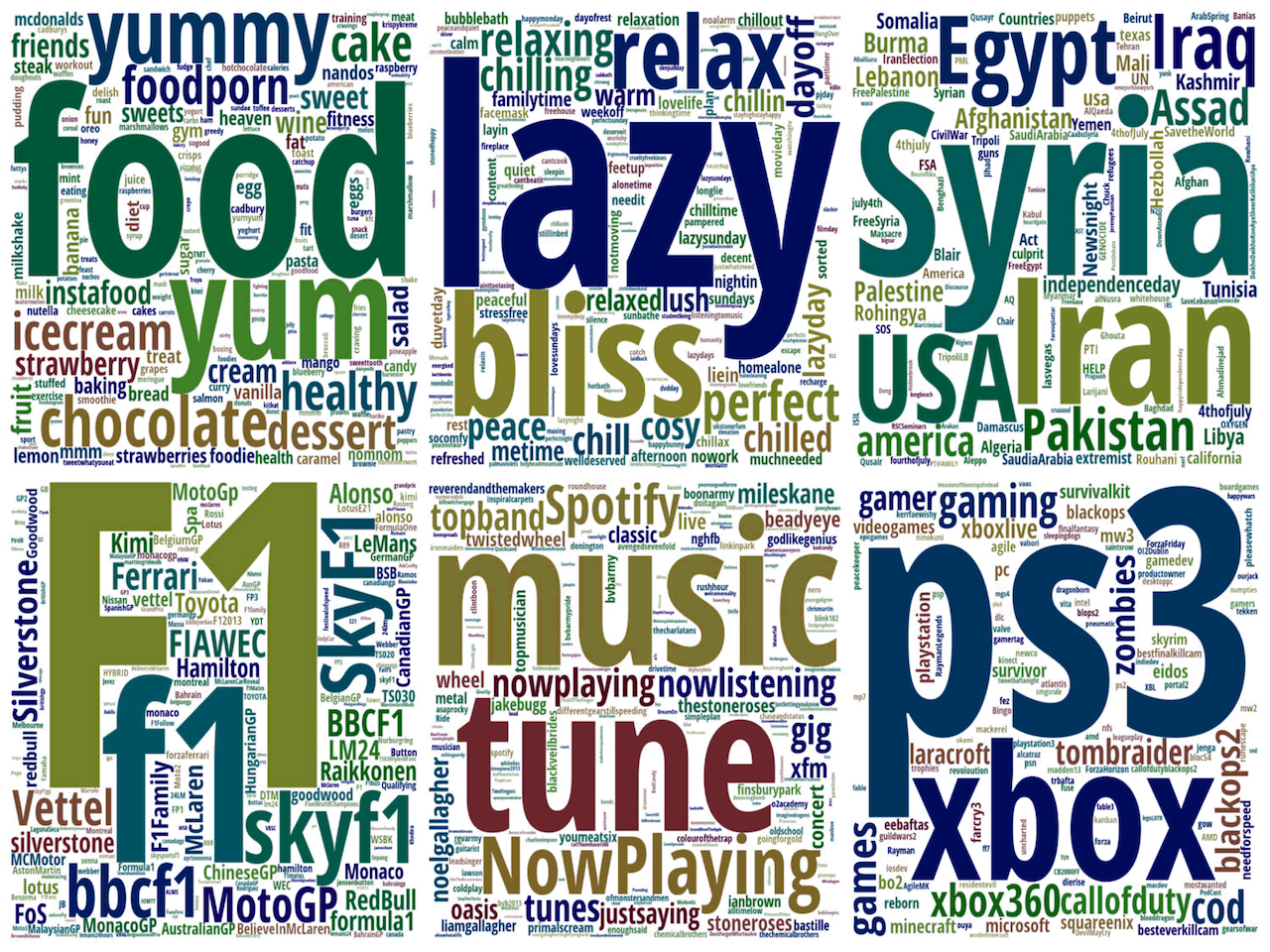}
	\caption{\textbf{Topics.}
		Word clouds for hashtags of six selected topics we detected in our Twitter data set. Hashtags' size in each figure is proportional to their degree in the co-occurrence subgraph of their community. See the text for further details. \label{fig:topics}}
\end{figure}

Fig \ref{fig:topics} shows examples of topics we detected in our Twitter data set. Users' affiliation to them indicates individual preferences in the wide range of topics available in the social network and it constitute our pool to assess similarity between users. The engagement in specific topics tells something about a user, and we adopt them as the basis to create a metric based on users' interests in different topics. In short, we want to assess if connected users tend to be topically similar and how much the similarity is relevant to their relationships.

Our results show that, on Twitter, \textit{follow} and \textit{mention} relationships are more likely to have a higher topical alignment than random pairs of users. Furthermore, we verify that both kinds of relationships tend to display a similar alignment pattern, despite the belief that they are relationships of a different kind \cite{Gonzalez-Bailon2014}. Finally, our analysis also shows that connections with strong interactions tend to have higher similarity and that the similarity between connected users indicates a higher probability of interaction. 

\section{Related Work} \label{sec:related}

In an online social system, the emergence of connections among individuals can be explained by different mechanisms from the preferential attachment\cite{Barabasi1999} to shortcuts for the consumption of information \cite{Weng2013a}. It is clear that the information shared in an online social network is an important characteristic to be taken into account while analyzing its connections.  However, there is no clear definition of information in a social network context. In this work, we consider information as the different kinds of content that flow in a network and may affect people's opinions or ideas. This is analogous to the Bateson's general definition of information as composed of pieces that are supposed to be \textit{``a difference that makes a difference"} \cite{Bateson1987}. Some recent efforts have been directed to the study of how the information traversing the network is related to its links. Weng et al. \cite{Weng2013a} recently demonstrated that information flows play an important role in link creation in the Yahoo! Meme network. Around 12\% of the new edges were motivated by the information flow, indicating that the network's edges dynamics cannot be explained merely by its topological structure. Furthermore, they showed that, while some users create connections mostly based on friendship, others are more guided by the content that users produce and share. 
Bogdanov et al. provide a model of pre-specified topics and verified the consistency of their use by Twitter users, they also applied this to predict influencers and to minimize the latency in information dissemination \cite{Bogdanov2013}.
Meyers et al. \cite{Myers2014} were interested in how the rise of abrupt changes in the information flow dynamics influences the creation and removal of links. Their work found that cascade of tweets was likely to cause \textit{follow} or \textit{unfollow} bursts, i.e., people start to follow or unfollow others with the abrupt increase in the retweets of some content.

Also using data from Twitter, Das et al.\cite{das2016} studied how the difference in users intent affects the content of their messages and their propagation. Suh et al.\cite{sun2010} focused on which features increase the probability of a message to be retweeted finding that the presence of hashtags  (i.e. the presence of a context), along with other factors, favors the sharing.  Following on how context and interests shape information sharing, Wu et al.\cite{wu2011} categorized influential users in Twitter -- i.e. celebrities, media, and bloggers -- finding that usually users in the same category show common behaviors that differ from one category to another. Another contribution along this line is the one by Kang and Lerman\cite{Kang2017} where they studied how the position in the network and the engagement of users affect the information they receive. The authors found that more engaged users usually occupy bridge positions in the network and are exposed to more diverse and novel information with respect to less engaged ones. Finally, the role of network structure and access to information has also been studied by Aral and Van Alstyne\cite{aral2011} analyzing data from an executive recruitment firm.

\subsection{Topical Alignment}
We are concerned with the degree to which users are more topically aligned with their connections. This is closely related to the homophily concept \cite{McPherson2001, McPherson1987, Newman2003, Kossinets2009, Lazarsfeld1954}, i.e., the tendency of individuals to form dyads with people similar to them, which have implications for the final network structure \cite{Javarone2013a}. If the similarity between pairs of individuals induces them to form a tie, this tendency is called \textit{choice homophily}, otherwise, if it is just a result of the constraints in the opportunities of connections, \textit{induced homophily}. Both types might be necessary to explain levels of similarity encountered in dyads, as Kossinets et al. showed with dyads of a university community \cite{Kossinets2009}. Nonetheless, choice homophily requires assessing individuals preferences and often this is infeasible. Thus, the concepts of \textit{baseline homophily}, the expected similarity between random pairs of individuals, and \textit{inbreeding homophily}, the similarity of dyads that are above or under the baseline, introduced by McPherson \& Smith-Lovin \cite{McPherson2001} are often used in practical approaches \cite{Robinson2009}. 

Topically aligned dyads are not necessarily a result of homophily, as connected individuals also tend to become more similar to each other over time, what is know as \textit{social influence}, or social contagion \cite{Christakis2012, Aral2009, Shalizi2011}.  Social influence is an important ingredient for synthetic models such as the one proposed by Robert Axelrod \cite{Axelrod1997} and is also verified in social networks \cite{Crandall2008}. However, real data also show that some effects attributed to social contagion may be a result of homophily \cite{Aral2009}. Furthermore, the creation of dyads may be motivated by latent or by unknown characteristics as pointed by Shalizi et al., thus, it might be impossible to verify whether ties similarity is really the result of homophily or social influence \cite{Shalizi2011}. This infeasibility in disentangling both processes does not affect our work, as we are not interested in verifying which one is driving the similarity of the dyads. Our goal is to assess to which degree connected users are topically similar, independently of the generating mechanism.

Nonetheless, we consider that the works more related to our own were strictly interested in homophily in online networks. Laniado et al. \cite{Laniado2016} inspected gender homophily -- i.e., the prevalence of same-gender relationships --  in the Tuenti Spanish social network. They based their analysis on self-reported gender data and their results showed the presence of gender homophily in dyadic and triadic relationships. Aiello et al. \cite{Aiello2012} explored homophily in the context of tagging social networks (Flickr, Last.fm, and aNobii). In these networks, tags are used to classify resources -- a different usage than hashtags on Twitter. In their approach, tags employed by the users are used to compute their similarity, which quantifies their proximity in tags usage. They found that users topical similarity is related to their shortest path distance on the social graph and that it could predict some links on the graph. Crandall \cite{Crandall2008} explored homophily using datasets extracted from Wikipedia and LiveJournal -- article and blogging based networks -- and modeled users according to their articles editing history. Choudhury explored homophily over a set of demographic users characteristics and its relation to the structure of their ego-network, most importantly they showed that the presence of homophily concerning topical interests is independent of the ego network structure \cite{Choudhury11}.

Some of these works are more related to networks centered in some kind of digital artifact, e.g., image, article, etc. Twitter, however, is more centered on the information posted by its users. Furthermore, hashtags or other features, by themselves, are not sufficient to assess similarity among users as they do not fully capture the context of users' messages. Thus, despite their findings, these works leave aside the latent semantics in the information sharing. Others had to rely on an external tool or specific classification to measure users similarity \cite{Choudhury11, NBERw20681, Kang2012a}.  It is necessary to look at a higher granularity to capture the different kinds of content that users are engaged with, which we achieve using topics of information. To the best of our knowledge, no study has explored the topic in this way.  Thereby, our work contributes to the understanding of the nature of relationships in a social network exploring a component still hard to be manipulated: the different kinds of information that traverse the network. 

\section{Methods and  Data} \label{sec:methods}

\subsection{Twitter Dataset}\label{sec:tdata}
There is no clear definition of social media or online social network,  however, there is a general consensus that services like Twitter are instances of social media services \cite{Obar2015}.  Due to its microblogging nature, some consider Twitter also as a news media or an information network \cite{Myers2014a,Kwak2010}. This is an important feature as we are interested in the content shared by the users and their relationships. We explore both \textbf{mentions}, mentioning a user in a tweet, and \textbf{follow}, subscribing to receive other user tweets, relationships in this work. In our analyses, we explicitly decided not to include \textbf{retweets} as we are more interested in information created by the users than shared information. Explicitly creating a new tweet supposes a larger effort than retweeting one, thus we believe that this is a more reliable proxy of users real interests with respect to retweets. Moreover, not considering retweets has also the side effect of limiting the number of bots in our datasets. As suggested in \cite{Ferrara2016} the majority of contents produced by bots are retweets. So excluding them and users with only retweets should reduce the number of bots in our analyses. Finally, the last interaction form present nowadays in Twitter --\textbf{quoted} tweets-- was not present in $2013$ when we started our data collection. Thus, we do not consider quoted tweets in this work.      

\begin{table}
	\centering

	\begin{tabular}{c|c}
		\textbf{Data} & \textbf{Raw} \\ \hline
		Tweets & 98,506,315 \\ \hline
		Tweets with Hashtags & 16,935,625 \\ \hline
		Distinct Hashtags & 4,320,429  \\ \hline
		Users with Tweets & 1,286,816 \\ \hline
		Users with Hashtags & 774,596 \\ \hline
		Population set & 608,899\\ \hline
		Central Users set  & 9,490 \\ \hline
	\end{tabular}
	\caption{Summary of Crawled Data. For a further description of users activity, see the Supplementary Information \label{tb:rawdata}}
\end{table}

Our dataset is composed by all the geo-localized tweets -- tweets with valid GPS coordinates -- located in the United Kingdom and Ireland  in a 7 months period from January to September 2013 through the Twitter's Streaming API \footnote{https://dev.twitter.com/streaming/overview, accessed in September 2016}. Further, more tweets of the users with geo-localized tweets and their follow/friend connections were obtained through the REST public API \footnote{https://dev.twitter.com/rest/public, accessed in September 2016}. The decision of focusing only on geo-localized tweets, although could partially limit our results, has several advantages. Firstly, it guarantees that the vast majority of the tweets are in the same language. Secondly, it allows us to focus more on discussions between users and local events rather than global events that are more likely to be influenced by other media sources. 
The final dataset, excluding retweets has 98 million tweets from January 18th to September 2nd, 2013. 


\subsection{Topics of Information}\label{sec:topics}

Information in Twitter flows through tweets, which are short messages with a highly dynamic vocabulary, encumbering  traditional text clustering techniques. We decided to build topics of information considering the tweets with hashtags, as they are indicators of the tweet content. Hashtags are users generated annotations containing a shared meaning, similar to acronyms generated organically by a population \cite{Javarone2013} Furthermore, it is common for users to insert more than one hashtag in a tweet, and we exploit this aspect to build a semantic mapping of information in Twitter. We assume the existence of a semantic association between hashtags that co-occur in the same tweet. This is analogous to the assumption that words are semantically associated if they are likely to co-occur frequently \cite{Turney2010}. Thus, our method focuses only on the implicit semantics given by Twitter messages, i.e., it does not consider explicit semantics given by other sources. This semantic mapping is captured by a weighted co-occurrence graph of hashtags, which we built by extracting all pairs of hashtags that co-occurred in each tweet in our dataset. Therefore, in this graph, an edge $(h_i,h_j)$ indicates that the hashtags $h_i$ and $h_j$ co-occurred and, as the graph is weighted, $w({h_i, h_j})$ gives the number of different tweets in which they are both present. 

We built a hashtag weighted co-occurrence graph using the 16,935,625 tweets with hashtags belonging to our dataset. As we removed hashtags that did not co-occur with any other, the co-occurrence graph resulted in 2,090,971 from the total of 4,320,429 distinct hashtags. As noted before, the edges of this graph represent a semantic association between hashtags. In order to further restrict our analysis to cases in which the statistics is not very scarce, and to reduce possible noise coming from low co-occurrences which might not have a clear significant association, we additionally removed all the edges between pairs of hashtags that co-occurred in less than 3 tweets. This process produces our final co-occurrence graph, which includes 104,308 hashtags and 526,522 edges.

We consider that topics of information are sets of hashtags clustered together in the graph. Thus, we expect that they will reflect the higher level structures that emerge from the latent semantic association of hashtags, providing the different contexts to which messages refer to. It is natural to see that these clusters could be captured by a community detection method and we decided to use the OSLOM tool \cite{Lancichinetti2011}. OSLOM is able to capture overlapping communities, a desirable feature considering that one hashtag may be used in different contexts.The application of OSLOM resulted in 2,074 communities and 14,118 homeless nodes, i.e., hashtags that did not belong to any community. We considered the communities and the homeless nodes as topics. Despite the latter possibly not significantly benefiting our future procedures, we believe that a hashtag alone can also carry information. Furthermore, our method to assess topical similarity should not be affected by this increase of topics as it does not take into consideration the topics that are not shared by two users (see the Supplementary Information for more details). Summing up both communities and homeless nodes in our analysis  we consider a total of 16,192 topics with an average of $622$ users per topic.

This approach of building a co-occurrence graph and using a community detection method to find topics was also used by Weng and Menczer \cite{Weng2015} through the Louvain method \cite{Blondel2008}, although they were not concerned with topical alignment. They assumed, based on the topical locality assumption, that semantically similar hashtags would appear in tweets together. Notwithstanding the resemblance to our premises, we do not presume that hashtags are similar, only semantically associated.  Even though there is not an easy way to ground the accuracy of this approach, we believe that it is a sound method for assessing information topics. Its premises and procedures are well defined over the semantic associations of hashtags. 

\subsection{Users Dataset}

Users considered in our analysis had, at least, one tweet with a hashtag in order to assess which topics of information they were affiliated with. Thus, we selected the 774,596 users from the 1 million of users with tweets. Before starting the analysis, we also took particular care in reducing the number of bots in our dataset. Along with excluding retweets we also decided to remove users that have been active for less than one day and those who showed an unusual activity. Specifically, we excluded users that had, on average, more than 400 tweets per day, as we consider that it is normally unfeasible for a real person to produce this quantity of tweets (for more information on bot filtering and their possible impact see the Supplementary Information).  Finally, users had to have, at least, one hashtag belonging to the topics detected (described in the previous section), leading to a final set of 608,899 users. We name this set \textit{Population} as it includes all the users in our experiment. 


After that, we extracted from the entire population another set of  $9490$ users, which we define as \textit{central users}. Those users are the core of our analyses as we calculate the topical similarity between them and their direct connections and compare it against random users selected from the entire population. Central users have been extracted randomly from the users in our dataset that have been active for the entire 7 months data collection period and produced at least $10^2$ tweets to guarantee a large corpus of tweets about their interests.
%
Details for the two sets are shown in Table \ref{tb:rawdata}. 

\subsection{Users Representation} \label{sub:usersrepre}

\begin{figure}[h!]
	\centering
	\includegraphics[width=0.7\columnwidth]{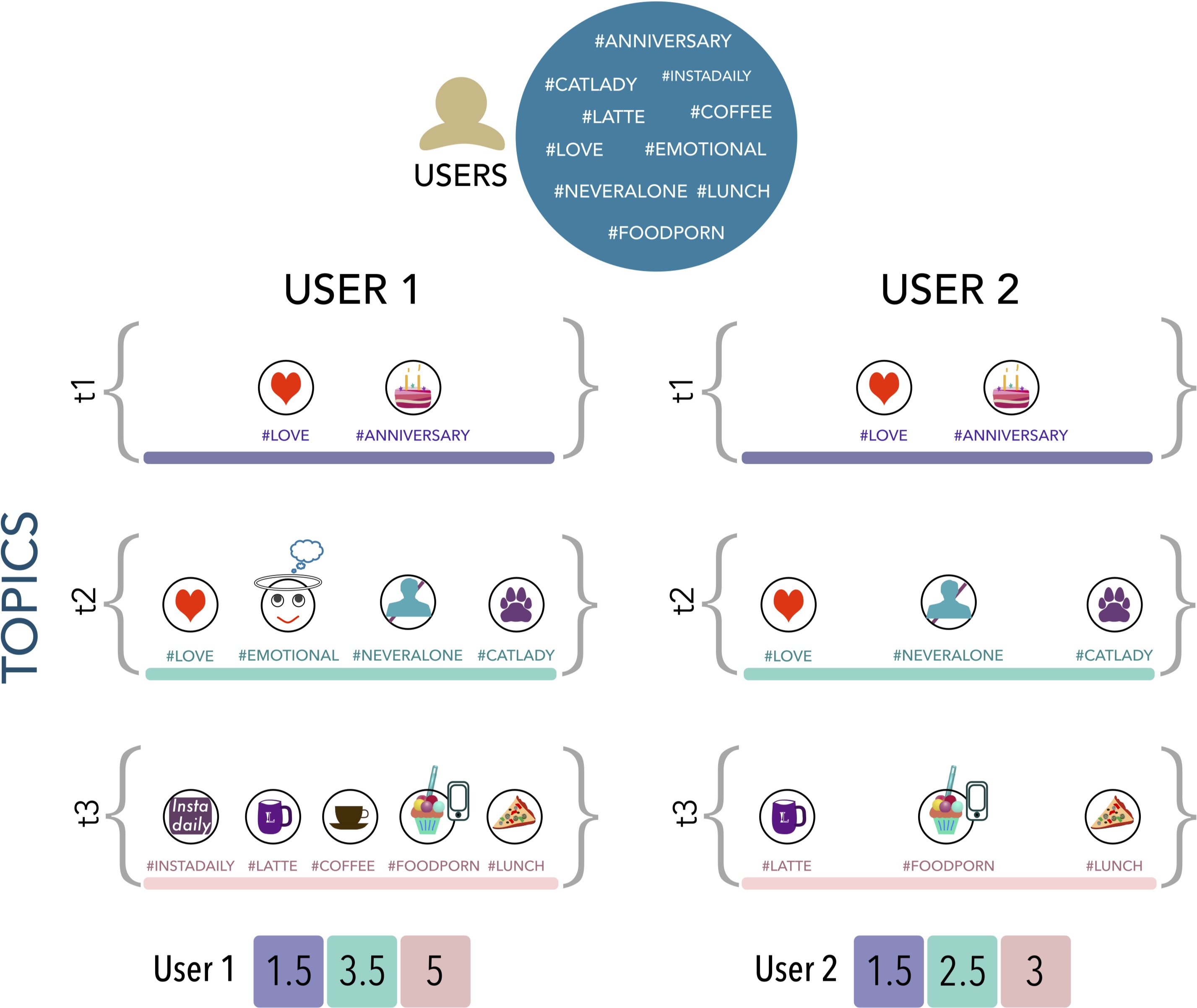}
	\caption{Extraction of a user feature vector from hashtags in different topics. The figure shows an example of how the feature vector is computed for two users that are part of the datasets. Hashtags shown are a small sample of those contained in the whole dataset and have been selected only to illustrate the example depicted.\label{fig:vector}}
\end{figure}

Each user is represented by a feature vector $\textbf{u}$, which comprises her affiliation to all topics of information. The process of building a user vector is illustrated in Fig \ref{fig:vector}. Feature $u_i$ corresponds to her affiliation in topic $i$ and its value represents the number of hashtags belonging to $t_i$(the set of hashtags belonging to the topic $i$) that were used by the user in her tweets. As the communities obtained by OSLOM may overlap, the same hashtag may be computed in more than one feature. In this case, each hashtag adds a proportional value to each feature it belongs . The value of a feature $u_i$ is given by 

\begin{equation} \label{eq:membership}
u_i \leftarrow \sum\limits_{\{h \in H : h \in t_i\}}{\dfrac {m_U(h)}{|\{t \in T: h \in t\}|}}
\end{equation}

All the hashtags used by a user are contained in a multiset $U=(H,m_U)$, wherein $H$ is the set of used hashtags and $m_U$ gives the number of occurrences of each hashtag. $T$ is the set of topics, i.e., communities of hashtags. Strictly speaking, each element $t \in T$ stands for a topic and it is a set containing the hashtags inside one cluster built by the community detection method. Fig \ref{fig:vector} illustrates a user multiset and its transformation in the user feature vector via Eq.~\ref{eq:membership}. As \textit{\#love} appears in the topics $t_1$ and $t_2$, it adds $1$ to their respective features.    

\subsubsection{Weighting Users' Vectors}
The previous definition of users' features vector considers that all topics have the same weight, i.e., the values of the respective features are directly derived from the number of hashtags used. This may be not suitable for our task as some popular topics or of general use could be over-represented and thus should have a smaller weight. To overcome this distortion, we consider that topics shared by a large percentage of the users ought to have a small weight, likewise, topics possessed by only a small percentage of users ought to weight more. The intuition behind this is that features corresponding to rare topics should be more discriminative of the topical proximity of users than features corresponding to frequent topics. 

Strictly speaking, we would like to take into account the information content of each topic \cite{Cover2006}. To do so, we rely on TF-IDF\cite{Turney2010} to weight users affiliation to each topic $u_{i}$ following:

\begin{equation}\label{eq:tfidf}
u_i \leftarrow u_i \times log \dfrac{|I|}{|\{v \in I: v_i >  0\}|}
\end{equation}

where $I$ is the set of all individuals, i.e., Twitter users. For each feature $i$ in the user vector, this method will weigh its value according to the number of users that also used it -- e.g., a feature that is shared by all users will have its value set to 0 as it does not provide information to discriminate users.

\subsection{Computing  Similarity between Users} \label{sub:similarity}

With the representation of users as feature vectors, we are able to compute topical similarity between two users using as metric the cosine similarity of their vectors \cite{Turney2010}. The cosine similarity fits well to this task as it only focuses on the angle between vectors -- i.e., it does not consider their length.  Cosine similarity ranges from 0 to 1; identical users would have similarity 1; users that do not share anything in common 0. It is evaluated using Eq.~\ref{eq:cos} below. In preliminary analyses, we also tested  Kendall's tau, Spearman's rho and Jaccard similarity measures. We did not adopt them as they did not present significant differences or improvements with respect to cosine similarity.
\begin{center}
\begin{equation}
\label{eq:cos}
sim_{cos}(u,v) = \dfrac{u \cdot v}{||u||||v||}
\end{equation}
\end{center}

\section{Topical Alignment} \label{sec:homophily}

The hypothesis that users are more topically aligned to their neighbors than to random users will be addressed here in terms of \textit{baseline alignment} and \textit{inbreeding alignment} similar to the classification introduced by McPherson \& Smith-Lovin \cite{McPherson2001}. Here, we consider \textbf{baseline alignment} as the expected average similarity between users and a random group of the population. Inbreeding alignment is defined as the difference between the baseline distribution and the distribution of average similarity between the users and those with whom they form a dyad, which is formed by a \textit{follow} or \textit{mention} relationship. In other words, baseline alignment is our null model and inbreeding alignment a measure of how much real values deviate from the null model. This deviation is captured by the Kolmogorov-Smirnov test \cite{Conover1999} and the likelihood of the distribution of dyads yielding higher (or lower) values of average similarity is captured by a Mann-Whitney U test \cite{Rice1995, McGraw}. We believe this approach has significant benefits than just looking at the hashtags shared by users as we comment on the Supplementary Information material.

\subsection{Topically Aligned Follow Relationships}

We initially explore inbreeding alignment with respect to \textit{follow} connections. Our hypothesis is that users are, on average, more similar with their followees, i.e., we expect their topical alignment to be significant. This means that the distribution of similarity averages of the individuals with their followees is expected to yield higher values than the distribution of averages with randomly chosen individuals from the population. We tested this hypothesis using the central users and their followees.

\begin{figure}[h!]
	\centering
	\includegraphics[width=0.85\columnwidth]{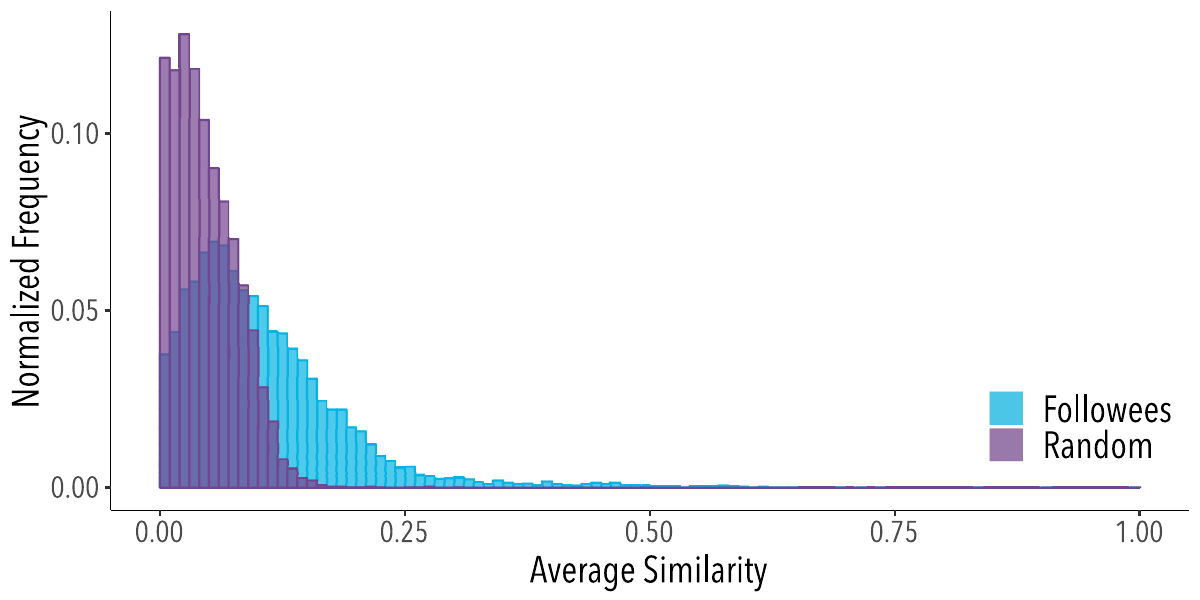}
	\caption{\label{fig:similarities}  Distribution of average similarity between central users and their followees (blue) and between central users and randomly selected groups of the same size (purple). KS(Kolmogorov-Smirnov statistic) = $0.37$, $p <0.001$, MW(Mann-Whitney U effect size) = $0.75$, $p <0.001$ The medians of the distributions of \textit{Followees} and the \textit{Random} are $0.087$ and $0.041$ respectively. Distributions have been calculated considering the whole set of 9,490 central users.}
\end{figure}

Fig \ref{fig:similarities} shows the histograms of the two distributions: \textit{Followees}, the distribution of averages computed for each central user with her followees; and \textit{Random}, the distribution wherein, for each central user, averages have been computed with a group composed of randomly chosen users from the \textit{Population} set, with the same size as the set of central user followees. As it can be seen, all the distributions are centered around low values of the cosine similarity spectrum. We consider that this effect is a result of the large number of topics and does not impact our results.

There is an overlap among the distributions, mostly concentrated in lower similarities. However, it is clear that there is a difference between the random distribution and the followees distribution.  The Kolmogorov-Smirnov statistics between the distributions  is $0.37$, $p <0.001$. We also used the Mann-Whitney U test to verify if the distribution with followees was likely to have a higher average similarity than the other. Results were positive with an effect size of $0.75$, $p <0.001$. Overall, the analysis shows that, on average, users tend to be connected to whom they are more similar with, that is, the similarity between followees is higher than the baseline similarity, thus showing the presence of inbreeding alignment. This implies that a user tends to have a stronger topical similarity with followees than with randomly chosen users.

\subsection{Users Interactions}\label{sec:inter}

\begin{figure}[h!]
	\centering
	\includegraphics[width=0.85\columnwidth]{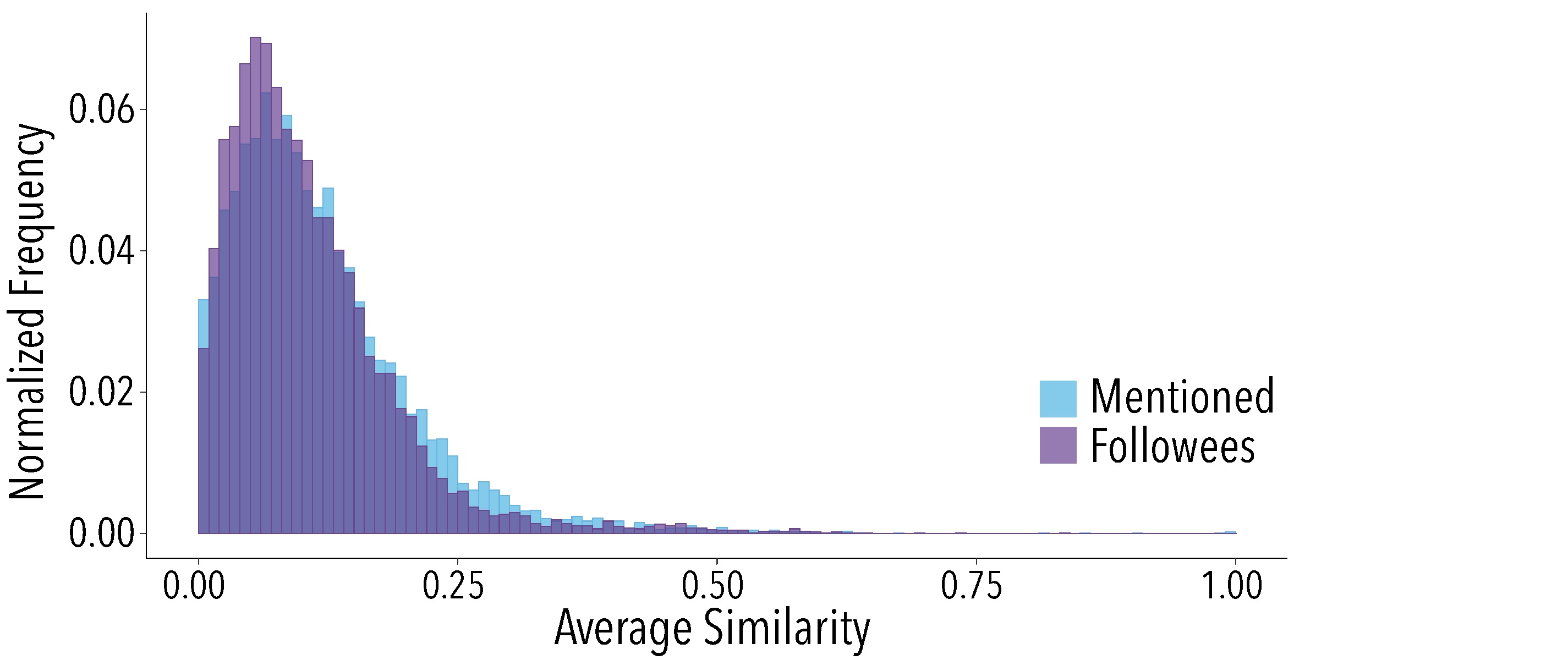}
	\caption{Distributions of average similarity between users followed (purple) and mentioned (blue) by central users. KS = $0.06$, $p <0.001$. Distributions have been calculated considering the whole set of 9,490 central users.\label{fig:histmentions}}
\end{figure}

Users on Twitter can use the convention \textit{@username} to mention another user in a tweet. The interactions that happen through mentions are often seen as a relationship stronger than the \textit{follow} connections \cite{Romero2011}. One hypothesis that emerges from such affirmation is that the topical similarity between mentioned users tends to be higher than between followed users. To test this hypothesis, we verified if the distribution of similarity averages with the mentioned users tended to be concentrated in higher values of similarity than the same distribution for followees. As shown by Fig \ref{fig:histmentions}, the distributions are roughly the same. Thus, in this context, we cannot say that the \textit{mention} relations are more topically aligned than the connections with followed users. 

\begin{figure}[h!]
	\centering
	\includegraphics[width=0.7\columnwidth]{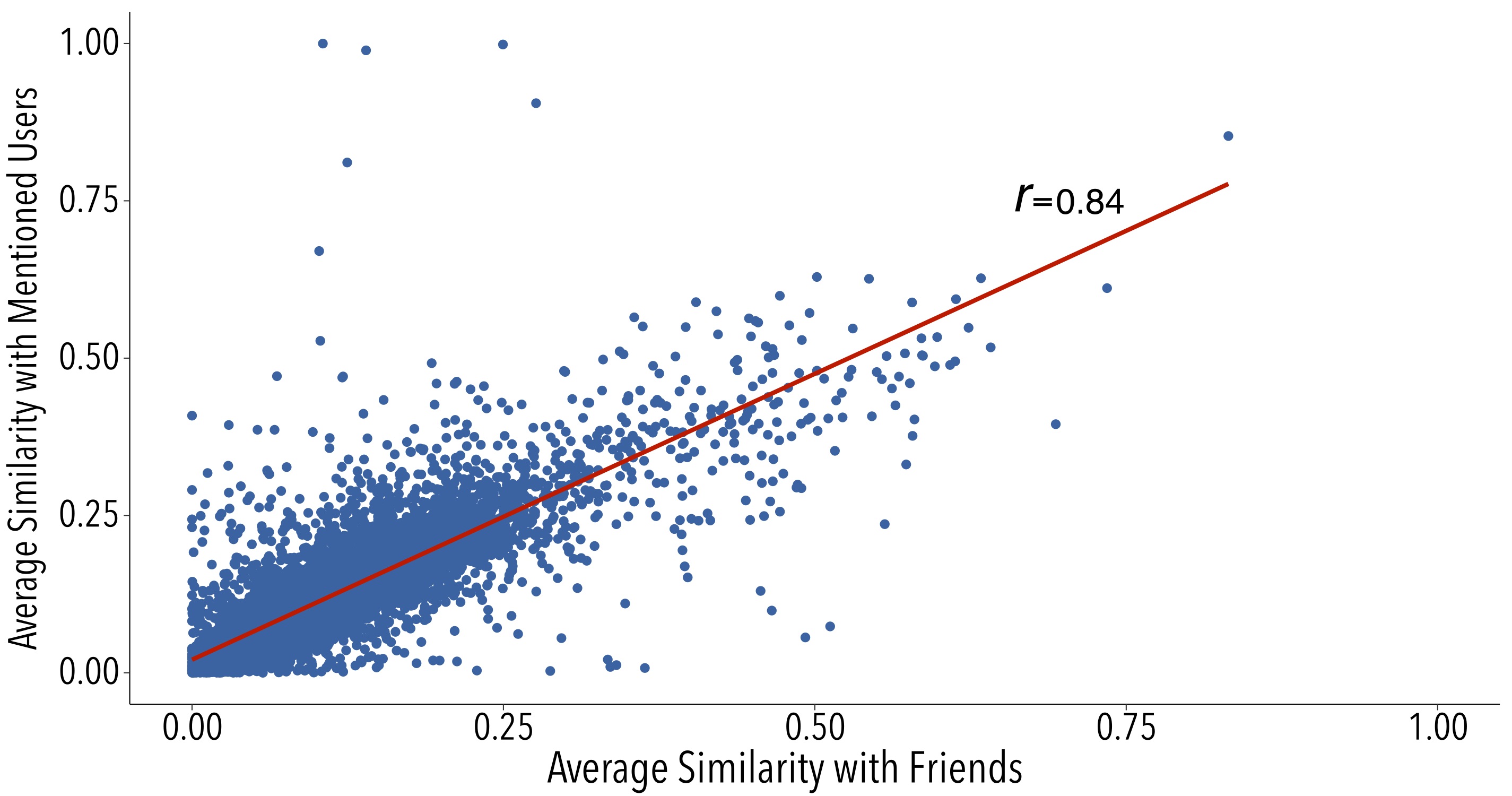}
	\caption{Correlation between average similarity with followees and mentioned users. Each point corresponds to the average similarity between a central user and the users she follows and the average similarity between the central user and the users mentioned by her. The Pearson correlation between the two variables is $0.84$.  \label{fig:allmentions}}
\end{figure}

Both \textit{mentions} and \textit{followees} histograms show that most of the averages fall into low values of similarity and there is a positive skewness of the two distributions, that is not evident in the distributions with random users  (Fig~\ref{fig:similarities}). Given the proximity between the two distributions presented in Fig \ref{fig:histmentions}, users on average might follow and mention others in a close similarity pattern. This hypothesis is verified in Fig \ref{fig:allmentions}, which indicates that users that tend to follow similar users, also tend to mention similar users.

\subsection{Reciprocity of Relationships} \label{sec:recstrength}

\begin{figure}[h!]
	\centering
	\includegraphics[width=1.0\columnwidth]{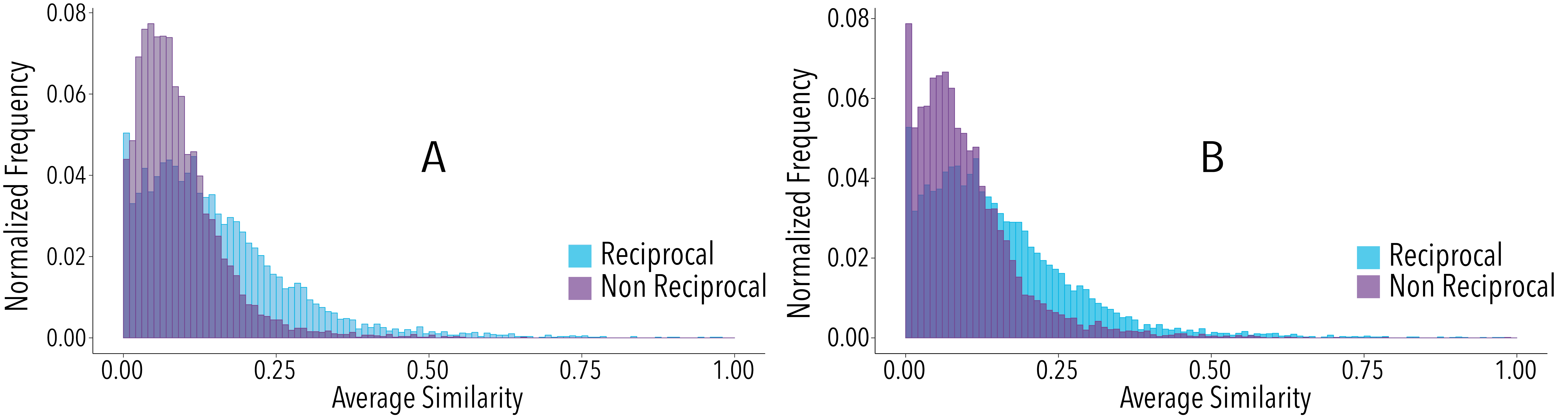}
	\caption{(\textbf{A}) Distribution of average similarity between central users and reciprocal (blue) and non reciprocal (purple) followees. KS = $0.27$, $p<0.001$, MW = $0.66$, $p <0.001$.
		Medians of the distributions of reciprocal and of nonreciprocal relationships are $0.12$ and $0.07$, respectively.
		Distributions have been calculated considering only  the 5,872 central users that have both reciprocal and non reciprocal followees in the dataset.
		(\textbf{B}) Distribution of average similarity between central users and reciprocal (blue) and non reciprocal (purple) mentions. KS = $0.22$, $p <0.001$, MW = $0.64$, $p <0.001$. The median similarity of the distribution of nonreciprocal mentions is $0.08$ while for reciprocal mentions is $0.12$. Distributions have been calculated considering only the 8,663 central users that have both reciprocal and non reciprocal mentions in the dataset.
		\label{fig:rec}}
\end{figure}

Relationships in Twitter are not reciprocal, a user following another does not imply that the other will choose to follow back. Thus, the existence of reciprocity indicates a stronger relationship between two users as both decided to establish this bond. In the scope of this work, the relationship strength is also viewed in terms of the topical similarity, thus, we expect that reciprocal dyads have a higher similarity than non-reciprocal dyads. This was verified for both \textit{mention} and \textit{follow} relationships, i.e., relationships wherein the two users mentioned each other and relationships in which the two follow each other. We first present the result regarding the reciprocity of the \textit{follow} connections in Fig \ref{fig:rec} (\textbf{A}). The two distributions differ, the distribution of similarity for the reciprocal followees is concentrated around higher values of similarity. The comparison for the reciprocal mentions distribution is shown in Fig \ref{fig:rec} (\textbf{B}).  The distribution of reciprocal mentions also has a higher similarity. This indicates that reciprocal relations are more prone to have a higher topical similarity, i.e., users have a more similar topic affiliation if they have a reciprocal relationship.

The tests conducted in this subsection reinforce what was seen in the previous section: there is no significant difference between the nature of \textit{mention} and \textit{follow} relationships with respect to topical similarity. The distributions of both relationships are very alike when considering the dyads similarity, even with reciprocal relationships. Furthermore, we could verify that, in the case of reciprocal relationships, there is a higher topical alignment than with nonreciprocal relationships. This indicates that users with a reciprocal relationship tend to become more similar by social influence or, conversely, that users similarity can be a factor which influences both to establish the relationship. Our method is unable to discriminate between either of the two mechanisms, as we would need to add a temporal dimension to the evolution of similarity and the network structure.


\subsection{Mention Probability} \label{sec:mentionprob}
All the analyses shown until now indicate that the similarity of most of the dyads is concentrated around low values. Therefore, it is natural to presume that most of the mentions made by central users involve users with low similarity with them. However, this contrasts with common sense as we expect that users in dyads with high similarity are more likely to be mentioned.

We explored this question, i.e., if the probability of being mentioned is higher for users with a high similarity, by looking at all dyads of followees. We also took into account the number of times that each followee was mentioned by a central user. To do so, we first defined $m_{u,v}$ as the number of mentions made by central user $u$ to followee $v$ and $s_{u,v}$ as their similarity. Then we calculated  $P(m_{u,v} > M | s_{u,v} \leq S)$ as the conditional probability of a user being mentioned more than $M$ times, given that her similarity with the mentioning user is smaller than $S$:

\begin{equation}
	\centering
\label{eq:mention_by_sim}
P(m_{u,v} > M | s_{u,v} \leq S) = \dfrac{P(m_{u,v} > M \cap s_{u,v} \leq S)}{P(s_{u,v} \leq S)}
\end{equation}

\begin{figure}[h!]
	\centering
	\includegraphics[width=0.7\columnwidth]{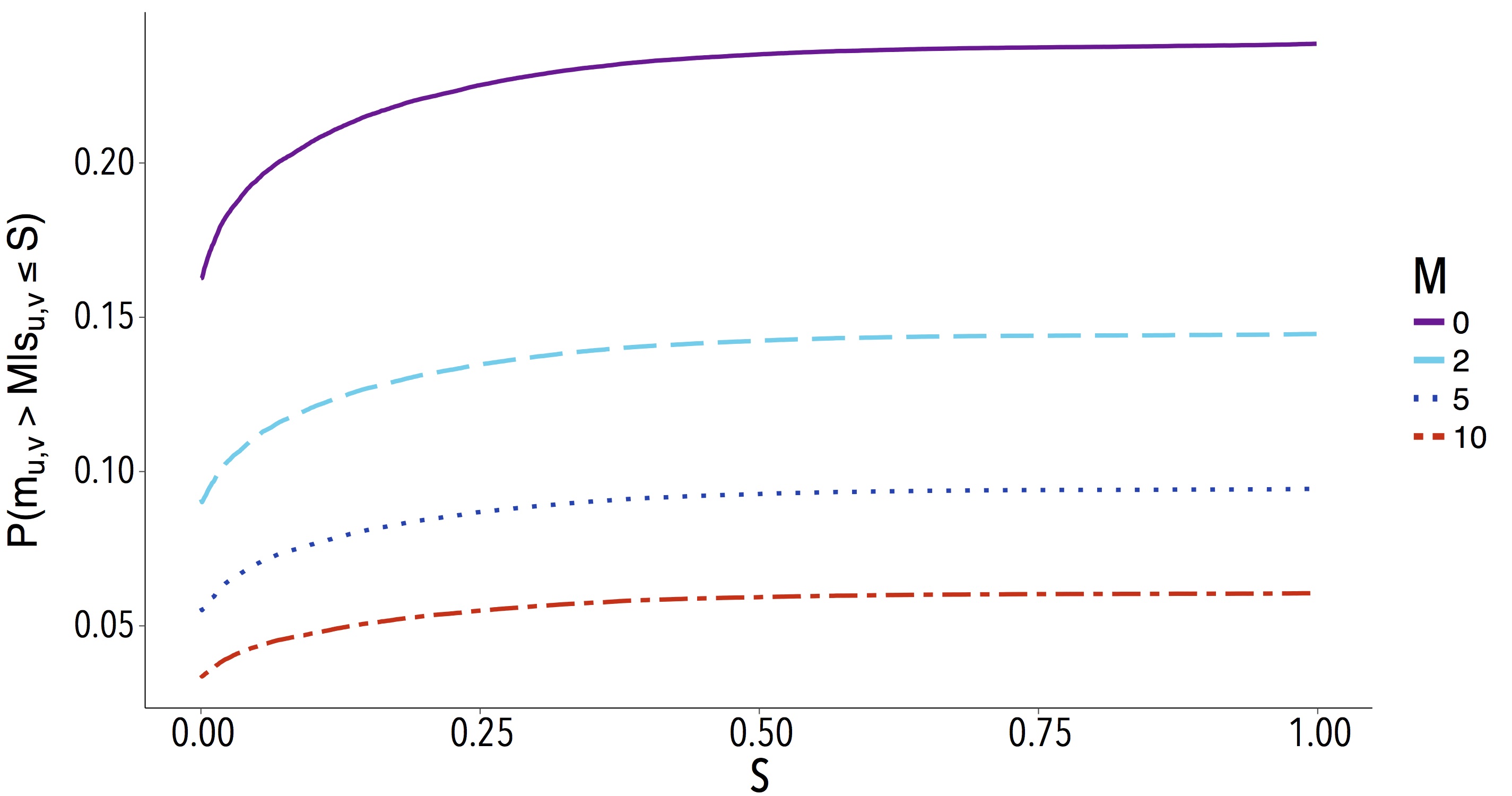}
	\caption{Conditional probability of followees being mentioned more than $M$ times by central users, given their similarity is smaller than $S$. The probability has been calculated using  $547,346$ dyads involving connected users. }
	\label{fig:cond_prob}
\end{figure}

Fig \ref{fig:cond_prob} shows the cumulative conditional probabilities of followees being mentioned by central users more than $M=$ 0, 2, 5 and 10 times, given their similarity with the central user. As expected, the probability decreases when the minimum number of mentions increases. Fig \ref{fig:cond_prob} also shows that followees which have low similarity with  central users do not have a higher probability of being mentioned.  Actually, it is observed a stable growth until  0.4 and, after that, all the curves  reach a plateau. Overall, the pattern of conditional probabilities appears to be the same for larger values of $M$, there is only a shift in the probability, as being mentioned more times is more challenging.

This analysis shows how the similarity gives an indication of the interactions inside connections, at least for some values of similarity. As more similar is the connected users, the higher is their probability to have interacted.

\subsection{Inference by Similarity}\label{sec:predictor}

There is a correlation between users average similarity with followees and mentioned users. It indicates that users, on average, follow and mention other users in a similar fashion with respect to topical similarity. However, until now, we did not provide a way to verify to which degree similarity among users is an indicator of their connections. In other words, we would like to know if topical similarity might be an effective way to predict relationships between users. Our question here is the following:  is it possible to infer a user's followees or mentions from a group of randomly selected users looking only at the similarity between them? 

Using pools of users of different sizes, we try to extract  from them all the connections of a central user considering their relative similarities. In this case, a pool always contains all the user's followees user mixed with other randomly selected users from the entire population. To create pools of different sizes we use a multiplicative factor $k$. The size of a pool is given by $k \times |fr(u)|$ where $fr(u)$ is the set of followees of user $u$ and $|fr(u)|$ is its cardinality. Thus, with $k=1$ the pool only contains $u$'s followees; for $k=2$ the pool will be constituted by all of $u$'s followees and the same number of random users. With $k=3$ we have all $u$'s followees and twice random users and so on until we reached pools of $60$ or $80$ times the size of the original set.

Once we created the pools, the similarity between central user  $u$ and all the users in the pool is computed. After that, a set with the same size of $fr(u)$ and containing the users that were most similar to $u$ is returned. Finally, this set is compared with the original set of followees of user $u$. To quantify the effectiveness of this method we calculated the average PPV(positive predictive values) for all the central users, i.e., the average of the fraction of followees that were correctly predicted. As previously mentioned, there are differences between users' average similarity, that suggest the presence of different following patterns. Thus, we repeated our analysis for users with different values of similarity, e.g. 0, 0.2, 0.4 and 0.6, along with considering all the central users together.  

\begin{figure}[h!]
	\centering
	\includegraphics[width=0.7\columnwidth]{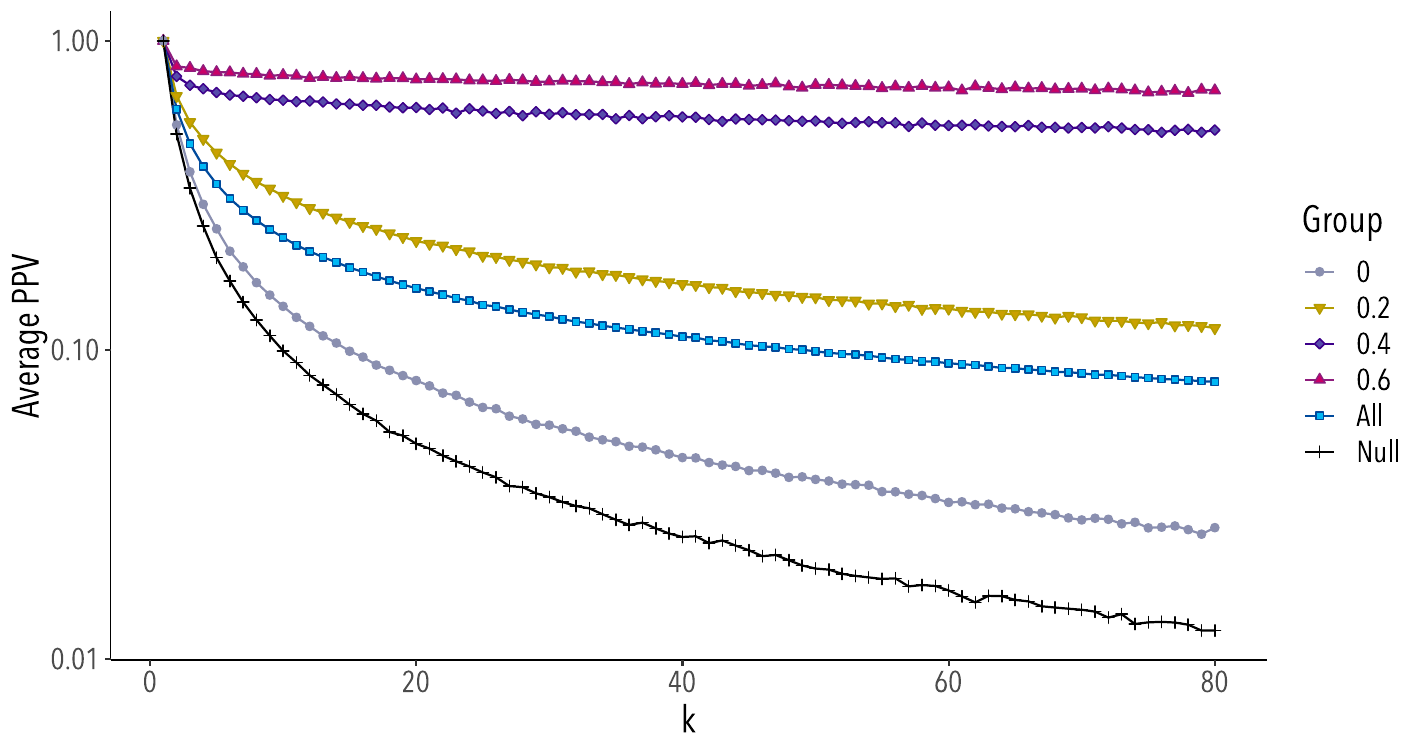}
	\caption{Average PPV(positive predictive values) of the inference mechanism for following connections with a pool of size $fr(u) \times k$ for different values of average similarity. The \textit{All} curve presents all the central users irrespectively of their average similarity while the \textit{Null} represents sets of randomly selected users.\label{fig:predictor_f}}
\end{figure}

Results are shown in Fig \ref{fig:predictor_f}. Each line shows the averages for each group of users. The blue line shows the average PPV considering all users together. To quantify the performance of this method against random selection, the red curve represents the average PPV if users were selected randomly instead of resting on similarity. For all the groups, our method outperforms random selection, indicating that similarity is an important feature in users connection process.

Results for an average similarity of $0.4$ and $0.6$ are worthy of a deeper analysis as the PPV remains roughly constant (or declines very slowly) for a wide range of values of $k$. This plateau in PPV means that even with an increasing set of users to choose from, the method keeps returning a significant fraction of their followees. This happens because they continue to be the most similar available in the whole pool. We believe that this is due to the fact that topics' affiliation patterns are almost unique for some dyads, hence, the majority of other users in the pool does not have a larger similarity than the actual followees of the user.

Even if the results for an average similarity of $0.4$ and $0.6$ are quite remarkable in terms of the match between inferred and real followees, the results obtained considering all the users together is not too good. Nonetheless, it is important to notice that the method applied here does not take into consideration the whole social network structure, which is likely the main factor responsible for determining connections. Our focus is to explore the relation between information and users' relationships, not to provide a complete algorithm for link prediction or recommendation. Having said that, we, however, believe that our results show that users affiliation in topics can be an important feature to be taken into account in link prediction or recommendation algorithms.

\begin{figure}[h!]
	\centering
	\includegraphics[width=0.7\columnwidth]{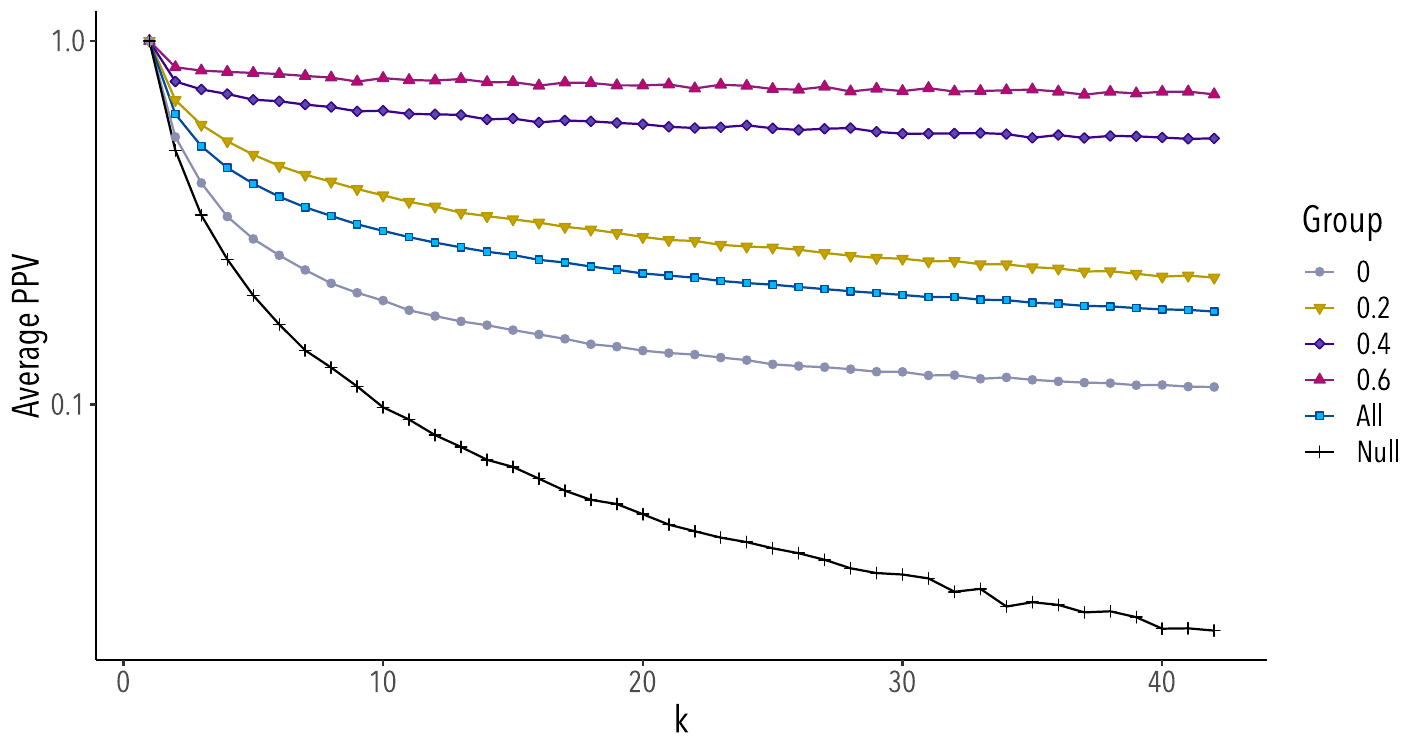}
	\caption{Average PPV(positive predictive values) of the inference mechanism for mentioning relations with a pool of size $fr(u) \times k$ for different values of average similarity. The \textit{All} curve presents all the central users irrespectively of their average similarity while the \textit{Null} represents sets of randomly selected users.\label{fig:predictor_m}}
\end{figure}

We repeated the process done for following relations considering, in this case, the probability of mentioning another user. In this case, we verified whether we could infer if a central user mentioned another user only looking at the similarity between them. Results are shown in Fig~\ref{fig:predictor_m} and are quite similar to the ones for the following probability with, in some cases, a better performance. This once again reinforces the idea that, in the case of topical alignment, following and mentioning interactions show a similar behavior and highlights the importance that topical similarity might have for some users.

\section{Conclusions} \label{sec:conclusion}
In today's world, online social networks as Twitter provide a laboratory where information and users connections are available for study. In this work, we analyzed how the pair-to-pair structure of a social network is related to the information shared on it. Connections in a social network are the substrate over which information flows, which makes their flow partially dictated by the network structure. However, information flow cannot be seen as an independent phenomenon; its contents can affect how individuals behave. For instance, people might be inclined to bond with others following the affinity in the information they share. On the other hand, information shared by an individual can make other users less prone to establish a bond with her. We have explored this relation using Twitter's information and connection data demonstrating that individuals which have a relationship tend to be more similar than expected regarding the information they share, i.e., connected users tend to be topically aligned.

On the other hand, in order to investigate how information is coupled with social connections, a key point is to design a model which captures its desired characteristics. We achieve this by modeling information as semantic topics of hashtags as Weng et al. \cite{Weng2015}. These topics encompass contents of information shared among users. We computed users affiliation in topics to characterize individuals' interests and preferences on Twitter. This characterization served as a basis for the exploration of topical similarity between individuals and we found that, on average,  individuals are more likely to have a relationship with more similar users. For some users this effect is so profound that they are essentially connected to the users most similar to them in all our dataset, which suggests an effective way to predict new connections at least for a subset of individuals in the network.

We have also verified if the influence of topical similarity between individuals differed in \textit{mentions} and \textit{follows} relations. Our results show a consistency across the two types of relationships, showing no significant difference between them. This was also verified when considering reciprocal relationships, which, in both cases, showed a higher level of similarity than non-reciprocal ones.

The approach presented in this work uses hashtags to build information topics. This limited our results to users that used hashtags, which significantly reduced our sample. Moreover, as we did not have the whole Twitter network structure, our hypothesis was restricted to exploring dyads and could not explore questions involving network measures, such as distance and centrality. Additionally, considering only geo-localized tweets further reduced the size of our datasets.  Nonetheless, we believe that our sample provides a significant support to understand some relationships among users.  There is also the possibility to improve our method to build topics, which currently ignores the temporal behavior of hashtags. The moment in which hashtags co-occur might contain specificities that we were not able to capture. However, even with these limitations, we could verify that the topics detected have a semantic sense and our datasets were sufficiently large as to achieve statistically relevance.

Our work demonstrates the importance of topical similarity between users regarding their connections and interactions. Our contribution also provides a feasible computational way to compute the similarity between users and can be used to further explore homophily and social influence in a social network. This can be further enhanced to improve our understanding of the mechanisms by which users connect, analyzing the whole social network structure, which was not available to us. Furthermore, it is necessary to further investigate how the flow of information is related to network dynamics. Our results also leave open opportunities to explore how topics' semantics affect the behavior of users who adopt them. Other possibilities include using our method in applications for link recommendation or finding missing links in social networks.

\section*{Competing interests}
The authors declare that they have no competing interests.
  
\section*{Data Availability Statement}
The data used in this study is available, in anonymized form, at DOI: 10.5281/zenodo.833390.

\section*{Data Ethics Statement}
All data used in this work have been obtained using the Twitter public API. We adhered to the Spanish Law for personal data protection, which does not require obtaining permission from an Ethical Committee to use public and anonymized Twitter data. We also confirmed that we followed Twitter's terms and conditions when conducting this study.

\section*{Author's contributions}
F.C., S.M., A.S. and Y.M. designed research; F.C. and S.M. performed research; F.C. analyzed data; F.C., S.M., and Y.M discussed the results; and F.C., S.M., A.S. and Y.M. wrote the paper
.
\section*{Funding}
FC and AS acknowledges support from Microsoft, Santander,
CAPES, CNPq and FAPESP Project 2015/01587-0.
S. M. acknowledges support from the Ram\'on y Cajal Program by MINECO, Spain.  Y. M. and S.M. acknowledge support from the Government of Arag\'on, Spain through 
a grant to the group FENOL, by MINECO and FEDER funds (grant FIS2017-87519-P) and by the European Commission FET-Proactive Project Multiplex (grant 317532). S.M. also acknowledge the Spanish State Research Agency, through the Mar\'ia de Maeztu Program for Units of Excellence in R\&D (MDM-2017-0711).

\bibliographystyle{frontiersinHLTH&FPHY} 

\begin{thebibliography}{10}
	
	\bibitem{Myers2014a}
	Myers S, Sharma A, Gupta P, Lin J.
	\newblock {Information Network or Social Network? The Structure of the Twitter
		Follow Graph}.
	\newblock WWW'14 Companion. 2014; p. 493--498.
	\newblock doi:{10.1145/2567948.2576939}.
	
	\bibitem{Lehmann2012}
	Lehmann J, Gon{\c{c}}alves B, Ramasco JJ, Cattuto C.
	\newblock {Dynamical classes of collective attention in twitter}.
	\newblock In: Proceedings of the 21st international conference on World Wide
	Web - WWW '12. New York, New York, USA: ACM Press; 2012. p. 251.
	
	\bibitem{Lee2011}
	Lee K, Palsetia D, Narayanan R, Patwary MMA, Agrawal A, Choudhary A.
	\newblock {Twitter Trending Topic Classification}.
	\newblock In: 2011 IEEE 11th International Conference on Data Mining Workshops.
	IEEE; 2011. p. 251--258.
	
	\bibitem{Bogdanov2013}
	Bogdanov P, Busch M, Moehlis J, Singh AK, Szymanski BK.
	\newblock {The social media genome}.
	\newblock Proceedings of the 2013 IEEE/ACM International Conference on Advances
	in Social Networks Analysis and Mining - ASONAM '13. 2013; p. 236--242.
	\newblock doi:{10.1145/2492517.2492621}.
	
	\bibitem{Gonzalez-Bailon2014}
	González-Bailón S, Wang N, Rivero A, Borge-Holthoefer J, Moreno Y.
	\newblock {Assessing the bias in samples of large online networks}.
	\newblock Social Networks. 2014;38:16--27.
	\newblock doi:{10.1016/j.socnet.2014.01.004}.
	
	\bibitem{Barabasi1999}
	Barab{\'a}si AL, Albert R.
	\newblock {Emergence of Scaling in Random Networks}.
	\newblock Science. 1999;286(5439):509--512.
	\newblock doi:{10.1126/science.286.5439.509}.
	
	\bibitem{Weng2013a}
	Weng L, Ratkiewicz J, Perra N, Gon\c{c}alves B, Castillo C, Bonchi F, et~al.
	\newblock The Role of Information Diffusion in the Evolution of Social
	Networks.
	\newblock In: Proceedings of the 19th ACM SIGKDD International Conference on
	Knowledge Discovery and Data Mining. KDD '13. New York, NY, USA: ACM; 2013.
	p. 356--364.
	
	\bibitem{Bateson1987}
	Foxman D, Bateson G.
	\newblock {Steps to an Ecology of Mind}.
	\newblock The Western Political Quarterly. 1973;26(2):345.
	\newblock doi:{10.2307/446833}.
	
	\bibitem{Myers2014}
	Myers SA, Leskovec J.
	\newblock {The bursty dynamics of the Twitter information network}.
	\newblock In: Proceedings of the 23rd international conference on World wide
	web - WWW '14. New York, New York, USA: ACM Press; 2014. p. 913--924.
	
	\bibitem{das2016}
	Das A, Gollapudi S, Kiciman E, Varol O.
	\newblock Information Dissemination in Heterogeneous-intent Networks.
	\newblock In: Proceedings of the 8th ACM Conference on Web Science. WebSci '16.
	New York, NY, USA: ACM; 2016. p. 259--268.
	
	\bibitem{sun2010}
	Suh B, Hong L, Pirolli P, Chi EH.
	\newblock Want to be Retweeted? Large Scale Analytics on Factors Impacting
	Retweet in Twitter Network.
	\newblock In: 2010 IEEE Second International Conference on Social Computing;
	2010. p. 177--184.
	
	\bibitem{wu2011}
	Wu S, Hofman JM, Mason WA, Watts DJ.
	\newblock Who Says What to Whom on Twitter.
	\newblock In: Proceedings of the 20th International Conference on World Wide
	Web. WWW '11. New York, NY, USA: ACM; 2011. p. 705--714.
	
	\bibitem{Kang2017}
	Kang JH, Lerman K.
	\newblock Effort Mediates Access to Information in Online Social Networks.
	\newblock ACM Trans Web. 2017;11(1):3:1--3:19.
	\newblock doi:{10.1145/2990506}.
	
	\bibitem{aral2011}
	Aral S, Alstyne MV.
	\newblock The Diversity-Bandwidth Trade-off.
	\newblock American Journal of Sociology. 2011;117(1):90--171.
	
	\bibitem{McPherson2001}
	Mcpherson M, Smith-Lovin L, Cook J.
	\newblock {Birds of a Feather: Homophily in Social Networks}.
	\newblock Annual Review of Sociology. 2001;27(1):415--444.
	\newblock doi:{10.1146/annurev.soc.27.1.415}.
	
	\bibitem{McPherson1987}
	McPherson JM, Smith-Lovin L.
	\newblock {Homophily in Voluntary Organizations: Status Distance and the
		Composition of Face-to-Face Groups}.
	\newblock American Sociological Review. 1987;52(3):370.
	\newblock doi:{10.2307/2095356}.
	
	\bibitem{Newman2003}
	Newman MEJ.
	\newblock {The Structure and Function of Complex Networks}.
	\newblock SIAM Review. 2003;45(2):167--256.
	\newblock doi:{10.1137/S003614450342480}.
	
	\bibitem{Kossinets2009}
	Kossinets G, Watts DJ.
	\newblock {Origins of Homophily in an Evolving Social Network}.
	\newblock American Journal of Sociology. 2009;115(2):405--450.
	\newblock doi:{10.1086/599247}.
	
	\bibitem{Lazarsfeld1954}
	Lazarsfeld PF, Merton RK.
	\newblock {Friendship as a social process: A substantive and methodological
		analysis}.
	\newblock In: Berger M, Abel T, Page C, editors. Freedom and Control in Modern
	Society. New York: Van Nostrand; 1954. p. 18--66.
	
	\bibitem{Robinson2009}
	Robinson DT, Aikens L, Aikens L.
	\newblock {Homophily}.
	\newblock In: Encyclopedia of Group Processes {\{}{\&}{\}} Intergroup
	Relations. 2455 Teller Road, Thousand Oaks California 91320 United States:
	SAGE Publications, Inc.; 2009.
	
	\bibitem{Christakis2012}
	Christakis NA, Fowler JH.
	\newblock {Social Contagion Theory: Examining Dynamic Social Networks and Human
		Behavior}.
	\newblock Statistics in Medicine. 2013;32(4):556--577.
	\newblock doi:{10.1002/sim.5408}.
	
	\bibitem{Aral2009}
	Aral S, Muchnik L, Sundararajan A.
	\newblock {Distinguishing influence-based contagion from homophily-driven
		diffusion in dynamic networks.}
	\newblock Proceedings of the National Academy of Sciences of the United States
	of America. 2009;106(51):21544--9.
	\newblock doi:{10.1073/pnas.0908800106}.
	
	\bibitem{Shalizi2011}
	Shalizi CR, Thomas AC.
	\newblock {Homophily and Contagion Are Generically Confounded in Observational
		Social Network Studies}.
	\newblock Sociological Methods {\&} Research. 2011;40(2):211--239.
	\newblock doi:{10.1177/0049124111404820}.
	
	\bibitem{Axelrod1997}
	Axelrod R.
	\newblock {The Dissemination of Culture: A Model with Local Convergence and
		Global Polarization}.
	\newblock Journal of Conflict Resolution. 1997;41(2):203--226.
	\newblock doi:{10.1177/0022002797041002001}.
	
	\bibitem{Crandall2008}
	Crandall D, Cosley D, Huttenlocher D, Kleinberg J, Suri S.
	\newblock {Feedback effects between similarity and social influence in online
		communities}.
	\newblock In: Proc. KDD'08. New York, New York, USA: ACM Press; 2008. p. 160.
	
	\bibitem{Laniado2016}
	Laniado D, Volkovich Y, Kappler K, Kaltenbrunner A.
	\newblock {Gender homophily in online dyadic and triadic relationships}.
	\newblock EPJ Data Science. 2016;5(1):19.
	\newblock doi:{10.1140/epjds/s13688-016-0080-6}.
	
	\bibitem{Aiello2012}
	Aiello LM, Barrat A, Schifanella R, Cattuto C, Markines B, Menczer F.
	\newblock {Friendship prediction and homophily in social media}.
	\newblock ACM Transactions on the Web. 2012;6(2):1--33.
	\newblock doi:{10.1145/2180861.2180866}.
	
	\bibitem{Choudhury11}
	Choudhury MD.
	\newblock Tie Formation on Twitter: Homophily and Structure of Egocentric
	Networks.
	\newblock In: 2011 IEEE Third International Conference on Privacy, Security,
	Risk and Trust and 2011 IEEE Third International Conference on Social
	Computing; 2011. p. 465--470.
	
	\bibitem{NBERw20681}
	Halberstam Y, Knight B.
	\newblock Homophily, group size, and the diffusion of political information in
	social networks: Evidence from Twitter; 2016.
	
	\bibitem{Kang2012a}
	Kang J, Lerman K.
	\newblock {Using lists to measure homophily on twitter}.
	\newblock AAAI workshop on Intelligent techniques for web. 2012; p. 26--32.
	
	\bibitem{Obar2015}
	Obar JA, Wildman S.
	\newblock {Social media definition and the governance challenge: An
		introduction to the special issue}.
	\newblock Telecommunications Policy. 2015;39(9):745--750.
	\newblock doi:{10.1016/j.telpol.2015.07.014}.
	
	\bibitem{Kwak2010}
	Kwak H, Lee C, Park H, Moon S.
	\newblock What is Twitter, a Social Network or a News Media?
	\newblock In: Proceedings of the 19th International Conference on World Wide
	Web. WWW '10. New York, NY, USA: ACM; 2010. p. 591--600.
	
	\bibitem{Ferrara2016}
	Ferrara E, Varol O, Davis C, Menczer F, Flammini A.
	\newblock The Rise of Social Bots.
	\newblock Commun ACM. 2016;59(7):96--104.
	\newblock doi:{10.1145/2818717}.
	
	\bibitem{Turney2010}
	Turney PD, Pantel P.
	\newblock From Frequency to Meaning: Vector Space Models of Semantics.
	\newblock J Artif Int Res. 2010;37(1):141--188.
	
	\bibitem{Lancichinetti2011}
	Lancichinetti A, Radicchi F, Ramasco JJ, Fortunato S.
	\newblock {Finding statistically significant communities in networks.}
	\newblock PloS one. 2011;6(4):e18961.
	\newblock doi:{10.1371/journal.pone.0018961}.
	
	\bibitem{Weng2015}
	Weng L, Menczer F.
	\newblock {Topicality and Impact in Social Media: Diverse Messages, Focused
		Messengers}.
	\newblock PLOS ONE. 2015;10(2):e0118410.
	\newblock doi:{10.1371/journal.pone.0118410}.
	
	\bibitem{Blondel2008}
	Blondel VD, Guillaume JL, Lambiotte R, Lefebvre E.
	\newblock Fast unfolding of communities in large networks.
	\newblock Journal of Statistical Mechanics: Theory and Experiment.
	2008;2008(10):P10008.
	
	\bibitem{Cover2006}
	Cover TM, Thomas JA.
	\newblock Elements of Information Theory (Wiley Series in Telecommunications
	and Signal Processing).
	\newblock Hoboken, New Jersey, United States: Wiley-Interscience; 2006.
	
	\bibitem{Conover1999}
	Conover WJ.
	\newblock Practical nonparametric statistics.
	\newblock Wiley series in probability and mathematical statistics: Applied
	probability and statistics. Hoboken, New Jersey, United States: Wiley; 1980.
	
	\bibitem{Rice1995}
	Rice JA.
	\newblock Mathematical Statistics and Data Analysis.
	\newblock No. p. 3 in Advanced series. Boston, Massachusetts, United States:
	Cengage Learning; 2006.
	
	\bibitem{McGraw}
	McGraw KO, Wong SP.
	\newblock {A common language effect size statistic.}
	\newblock Psychological Bulletin. 1992;111(2):361--365.
	\newblock doi:{10.1037/0033-2909.111.2.361}.
	
	\bibitem{Romero2011}
	Romero DM, Meeder B, Kleinberg J.
	\newblock {Differences in the mechanics of information diffusion across
		topics}.
	\newblock In: Proceedings of the 20th international conference on World wide
	web - WWW '11. New York, New York, USA: ACM Press; 2011. p. 695.
	
	\bibitem{Javarone2013a}
	Javarone, MA and Armano, G.
	\newblock {Perception of similarity: A model for social network dynamics.}
	\newblock Journal of Physics A: Mathematical and Theoretical. 2013; 46(45):455102.

	
	\bibitem{Javarone2013}
	Javarone, MA and Armano, G.
	\newblock {Emergence of acronyms in a community of language users.}
	\newblock European Physical Journal B. 2013; 86(11):474.
	

	
\end{thebibliography}



\section*{Supplementary material}

# Supplementary Information: Topical alignment in online social systems

Felipe Maciel Cardoso[1,2,*], Sandro Meloni[2,4], André Santanchè[1], Yamir Moreno[2,3,5]

**1** Institute of Computing, State University of Campinas, Campinas, 13083-852, Brazil
**2** Institute for Biocomputation and Physics of Complex Systems, University of Zaragoza, Zaragoza, E-50018, Spain
**3** Department of Theoretical Physics, University of Zaragoza, Zaragoza, E-50009, Spain
**4** IFISC, Institute for Cross-Disciplinary Physics and Complex Systems (CSIC-UIB), 07122 Palma de Mallorca, Spain
**5** ISI Foundation, Turin, I-10126, Italy

\* fmacielcardoso@gmail.com

# 1 More information on Bots filtering

The presence of Bots in our dataset may alter the results of our analyses. Usually bots tend to produce a large number of tweets, mainly focused on one or few topics, so not filtering them could lead to large errors on the average similarity between users.

To assure that our dataset only contains a negligible fraction of Bots we implemented three different filtering strategies. First of all, as bots content tends to be dominated by retweets (i.e. see [1]) we decided to exclude all the retweets from our dataset and all the users with a very high percentage of retweets in their timeline. This decision has also been motivated by the fact that, in our analysis, we are more interested in contents produced by the users than in shared information.

To further reduce the probability of finding a bot in our dataset we also removed users with an unfeasible daily rate of tweets. 2 shows, for each user in our dataset, the average number of produced tweets per day in relation to the time they have been active; counted as the time difference between the first and the last tweet in our dataset. As it is clear from the figure, there is a high peak

in the activity of users with a short active time suggesting that they produced a high number of tweets and then disappeared. This is another typical signature of bots activity so we decided to filter all the users that have been active for one day or less and those who produced, on average, more than 400 tweets per day. This left 9490 central users and a total population of 608899 over the initial 774596 that used at least one hashtag included in the topics we extracted. The distribution of tweets per user is very heterogeneous, as it it can be seen on Fig. of this population can been in Fig. 1.

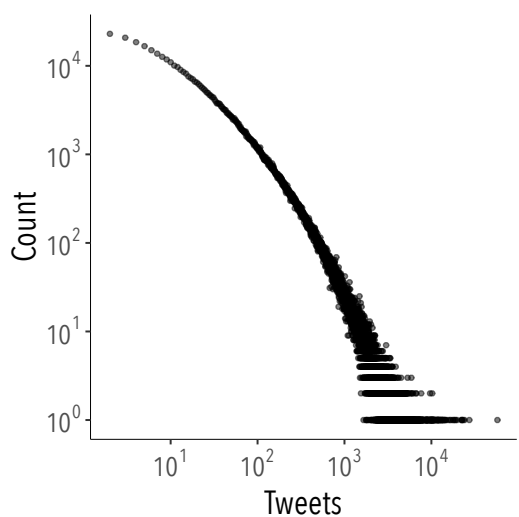

Figure 1: Distribution of tweets per user from the *Population* set.

Finally, we checked the quality of our filtering using the tool developed by Menzcer *et al.* – i.e. [1]–, Botomoter[1]. The tool, starting from a user ID, retrieves its timeline and returns a score between 0 and 1 that the user is a bot. A score of 1 means that almost certainly the user is a bot while 0 stands for a human user. Due to rate limits in the number of users that can be tested per day we decided to check only the central users in our dataset. 3 shows the distribution of scores of the central users with still an active account at the time of the analysis (6539 over the original 9490) while 4 presents the number of users with a score greater or smaller than 0.6. From these results, it is clear that the influence of bots on our analysis is minimal. The vast majority of the users in our dataset have a score lower than 0.25 and only 222 over 6539 have a score

---

[1]https://botometer.iuni.iu.edu/

higher than 0.6.

To further demonstrate the robustness of our analyses, we also computed the average similarities such as that on Fig 3 of the main text only considering central users whose score is smaller than 0.5 (5). Comparing the original figure with the new one, it is clear that, even using this conservative threshold, the possible influence of bots in our results is insignificant.

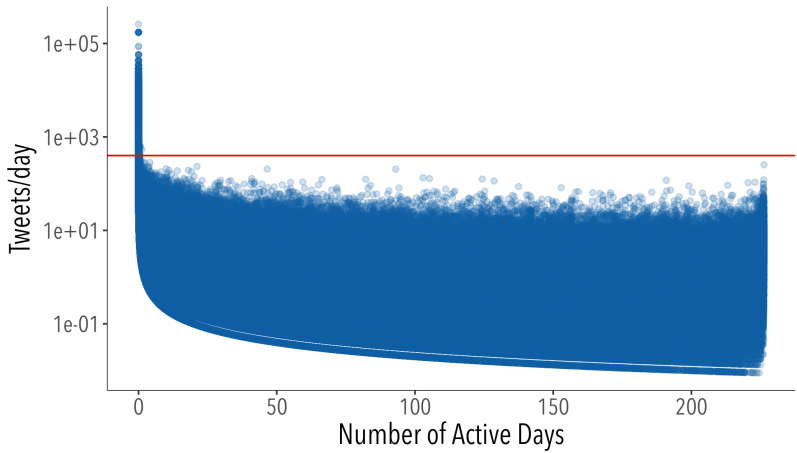

Figure 2: Average number of tweets per day generated by each user as function of the active time (in days) calculated as the time difference between the first and last tweet in our dataset.

## 2 Topical similarity versus hashtags similarity

One of the main innovations of our work is that we use topics instead of hashtags to calculate similarity between users. At this point one can argue what are the advantages of using topics instead of hashtags, as extracting topics from hashtags co-occurence network is a costly process. A similar approach could be considering directly vectors that describe hashtags usage instead of topics. This method, however, disregard dyads wherein users do not use the same hashtags but are interested in the same issues. To test if our method gives better results than only using hashtags we repeated the analysis in Fig 3 of the main text calculating the average similarity also using hashtags vectors. As demonstrated by 6, the distribution of average similarity calculated using hashtags is more peaked and centered at lower values. This is due, as shown in the inset of 6, to

3

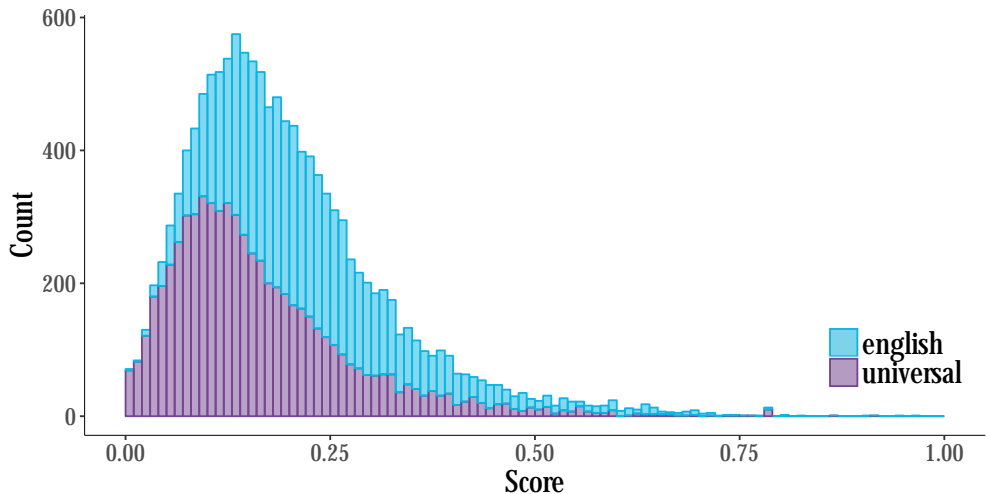

Figure 3: Bots score distribution for the central users of our dataset as given by the Botometer tool (https://botometer.iuni.iu.edu/).

the presence of a significant amount of dyads with a low similarity. This means that most connections use the same topic but not the same hashtags. Thus, we believe that using topics to detect similarity is more robust and allows to uncover relationships that would go unnoticed using hashtags.

## 3 The role of homeless nodes in topics detection

The extraction of topics based on community detection highlighted a high number (14118) of "homeless nodes" – hashtags that do not belong to any community, so the algorithm creates a new community with only one node –. Given their high number with respect to larger communities, it could be important to asses how those topics affect our results. Our intuition is that homeless hashtags are usually typos or ambiguous words not so common as hashtags. Thus, they should appear in few tweets and be employed by few users. To verify this hypothesis, Fig. 7 presents the number of homeless hashtags used by distinct users. As expected, more than half of the homeless hashtags have been used by only one user and almost the totality by no more than five. This supports our intuition that those hashtags play a minimal role in our results as topics used by only one user are not considered in the similarity.

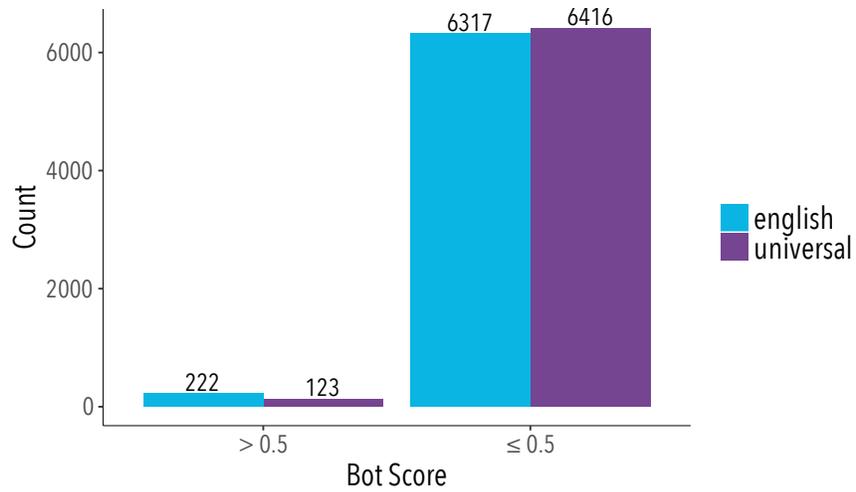

Figure 4: Number of central users classified as bots (score > 0.6) by the Botometer tool(https://botometer.iuni.iu.edu/).

## 4  A description of Topics

The majority of topics considered in this project contains few hashtags, as it can be seen in Fig. 8. The average number of hashtags without the homeless nodes is of 45.6 and of 6.7 taking the homeless nodes into account. Communities overlap and, on average, each hashtag belong to 1.04 communities. Communities are also quite heterogeneous with respect to their number of users, with an average of 622 users per topic.

## References

[1] Ferrara E, Varol O, Davis C, Menczer F, Flammini A. The Rise of Social Bots. Commun ACM. 2016;59(7):96–104. doi:10.1145/2818717.

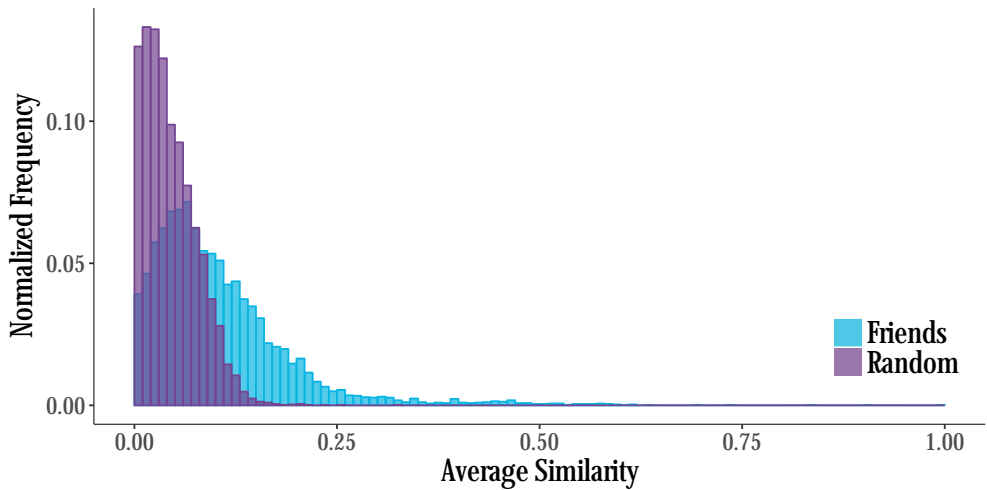

Figure 5: Distribution of average similarity between central users and their followees(purple) and between central users and randomly selected groups of the same size (yellow). Distributions have been calculated considering the whole set of 9,490 central users.

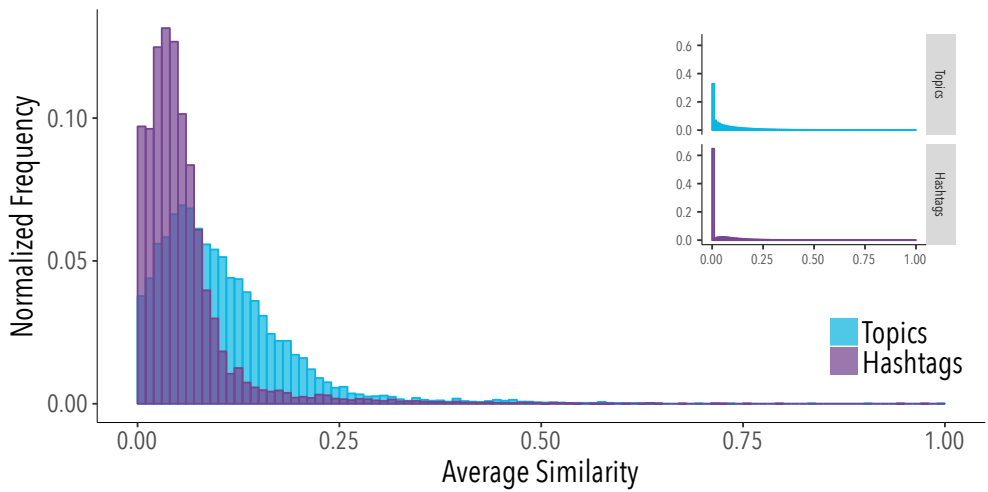

Figure 6: Distribution of average similarities of central users with their followees computed considering topics and hashtags. The inset shows the distribution of similarities of all the dyads.

6

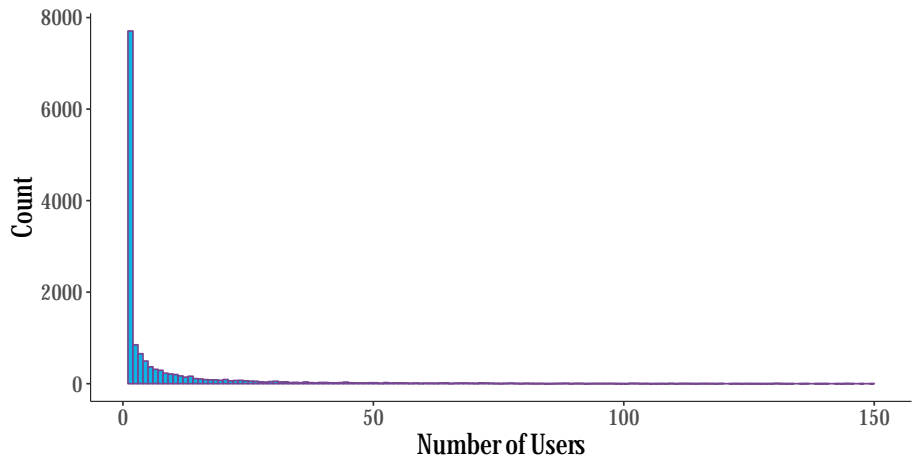

Figure 7: Number of the homeless hashtags used by distinct users.

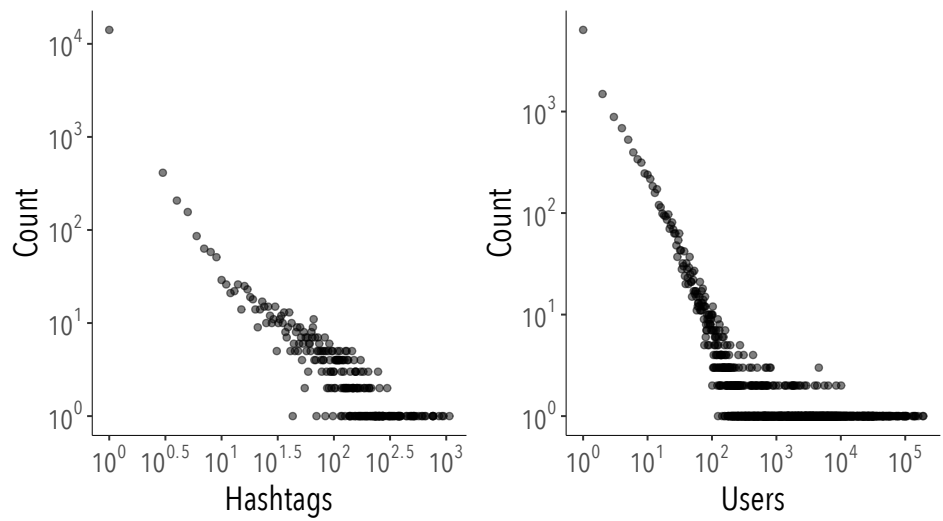

Figure 8: Distribution of hashtags (left) and users (right) of the topics.

7



\end{document}